# Frequency-Dependent Perceptual Quantisation for Visually Lossless Compression Applications

Lee Prangnell

Department of Computer Science, University of Warwick, England, UK

**Abstract** — The default quantisation algorithms in the state-of-the-art High Efficiency Video Coding (HEVC) standard, namely Uniform Reconstruction Quantisation (URQ) and Rate-Distortion Optimised Quantisation (RDOQ), do not take into account the perceptual relevance of individual transform coefficients. In this paper, a Frequency-Dependent Perceptual Quantisation (FDPQ) technique for HEVC is proposed. FDPQ exploits the well-established Modulation Transfer Function (MTF) characteristics of the linear transformation basis functions by taking into account the Euclidean distance of an AC transform coefficient from the DC coefficient. As such, in luma and chroma Cb and Cr Transform Blocks (TBs), FDPQ quantises more coarsely the least perceptually relevant transform coefficients (i.e., the high frequency AC coefficients). Conversely, FDPQ preserves the integrity of the DC coefficient and the very low frequency AC coefficients. Compared with RDOQ, which is the most widely used transform coefficient-level quantisation technique in video coding, FDPQ successfully achieves bitrate reductions of up to 41%. Furthermore, the subjective evaluations confirm that the FDPQ-coded video data is perceptually indistinguishable (i.e., visually lossless) from the raw video data for a given Quantisation Parameter (QP).

**1.0 Introduction**

The HEVC standard [1, 2] includes finite precision integer approximations of the Discrete Cosine Transform (DCT) and the Discrete Sine Transform (DST) [3, 4]. DCT and DST transform the intra prediction and inter prediction residual data from the spatiotemporal domain into the frequency domain. The DCT and DST basis functions correspond to the MTF characteristics of the Human Visual System (HVS). The DC transform coefficient and the low frequency AC transform coefficients contain the most important energy in terms of how a human observer perceives the reconstructed video data. Note that the primary objective of uniform scalar quantisation in HEVC is to discard perceptually irrelevant high frequency AC coefficients. This is because the HVS is not particularly sensitive to small gradations (i.e., imperceptible distortions) in quantised high frequency data. URQ is the default uniform quantisation technique in HEVC [5, 2, 4]; it quantises luma and chroma Cb and Cr transformed residual values (i.e., after the DCT and DST linear transformations). URQ applies equal levels of quantisation to all transform coefficients in luma and chroma TBs irrespective of the frequency sub-band in which a coefficient resides. The level of quantisation applied to a TB is determined by a Quantisation Step Size (QStep) value, which is controlled by an integer QP. URQ is not perceptually optimised, thus leaving room for improvement.

RDOQ [6, 7] is a more sophisticated approach to quantisation compared with URQ alone. The core objective with the RDOQ technique is to establish an optimal quantisation level for each transform coefficient in luma and chroma TBs. RDOQ measures the quantisation-induced distortion and also the number of bits required to encode the corresponding quantised transform coefficient. Based on these two values, the RDOQ chooses an optimal coefficient value, which is determined by finding an appropriate trade-off between bitrate and quantisation-induced distortion (known as the rate-distortion cost). RDOQ is widely adopted and, as such, it is the default quantisation technique utilised in both HEVC and H.264 (in combination with URQ). Due to the wide adoption of RDOQ in combination with URQ, Quantisation Matrices (QMs) are seldom utilised in contemporary HEVC applications. Although RDOQ can be considered as an advanced coefficient-level quantisation technique, it has been designed with a significant emphasis on improving mathematical reconstruction quality, as quantified by PSNR (compared with URQ and quantisation matrices, for example). Moreover, RDOQ is not perceptually optimised; it does not take into account HVS-related psychovisual redundancies.

The proposed FDPQ method is a perceptual coding scheme designed specifically for visually lossless compression in video coding applications. FDPQ is an MTF-based and perceptual compression variant of our previously proposed adaptive quantisation method in [8]. Whereas the technique in [8] increases the mathematical reconstruction quality of the compressed video data, FDPQ decreases video reconstruction quality in order to achieve considerable bitrate savings. Moreover, note that the experimental setup employed to evaluate FDPQ in this paper is also utilised in the unpublished thesis in [9]. FDPQ exploits the well-established MTF characteristics of the HVS as per the DCT and DST linear transformation basis functions utilised in the frequency domain (during the residual coding process). According to Parseval's theorem, the Euclidean distance between two points in a spatial domain is the same as the corresponding distance in the domain of a linear transformation (e.g., DCT). As such, by employing a Euclidean distance parameter in the frequency domain, FDPQ takes into account the distance of Y, Cb and Cr AC coefficients from the corresponding DC coefficient. In other words, FDPQ quantises more coarsely the least perceptually relevant transform coefficients (i.e., the high frequency AC coefficients). In addition, FDPQ preserves the integrity of the DC coefficient and thus facilitates a reduction of non-zero quantised AC coefficients. For this reason, the perceptually irrelevant quantised AC coefficients can be compressed more efficiently during the entropy coding process (i.e., via CABAC). The proposed method is compatible with raw video data of any bit depth and any YCbCr chroma sampling ratio. The objective evaluations reveal that FDPQ attains noteworthy bitrate reductions and, furthermore, the subjective evaluations show that the FDPQ-coded data is perceptually indistinguishable from the raw data (at *QP* = 17 and *QP* = 22) in all tests.

The rest of the paper is organised as follows. Section 2 includes background information relevant to the proposed method. In section 3, the proposed technique is presented. In section 4, thorough subjective and objective experimental evaluation results are presented and discussed. Finally, section 5 concludes the paper.

## 2.0 Related Background

To recapitulate, HEVC includes finite precision integer approximations of the DCT and the DST [3, 4]. These techniques transform intra prediction and inter prediction residual data from the spatiotemporal domain into the frequency domain. Recall that the DC transform coefficient and the low frequency AC transform coefficients contain the most important energy in terms of how the HVS perceives the reconstructed video data. As such, after intra prediction and/or inter prediction, DCT and DST are applied to the corresponding residual signals, from which transform coefficients are derived. More specifically, the DCT is applied to intra residual luma and chroma residual blocks of size 8×8 to 32×32. For inter-predicted residuals, the corresponding integer approximation of DCT is utilised on 4×4 to 32×32 luma and chroma residual blocks. Note that, for 4×4 intra residue, the DST is utilised instead of DCT. Recall that the integer DCT and DST schemes in HEVC exploit the MTF characteristics of the HVS. This is achieved by compacting the energy of luma and chroma prediction residual samples into the DC coefficient and the very low frequency AC coefficients.

As previously mentioned, it is well known that the DC coefficient and the low frequency AC transform coefficients are more important than the high frequency AC coefficients (in terms of how the reconstructed signal is perceived by the HVS). Because each coefficient frequency sub-band in a TB constitutes a different level of perceptual importance in a compressed video signal [10], the distance of AC coefficients from the DC coefficient can be quantified in terms of Euclidean distance. That is, the DC coefficient is the starting point and the distance of each AC coefficient from the DC coefficient represents the perceptual importance of the current AC coefficient.

Recall that URQ is the default uniform quantisation method in HEVC. Assuming that chroma QP offsets are not employed in HEVC, the QStep computation for luma data is identical to the QStep computations applied to chroma Cb and Cr data. As such, the luma and chroma QSteps in HEVC, denoted as $QStep_Y$, $QStep_{Cb}$ and $QStep_{Cr}$ are defined in (1) to (3), respectively:

$$QStep_Y(QP_Y) = 2^{\frac{QP_Y-4}{6}} \qquad (1)$$

$$QStep_{Cb}(QP_{Cb}) = 2^{\frac{QP_{Cb}-4}{6}} \qquad (2)$$

$$QStep_{Cr}(QP_{Cr}) = 2^{\frac{QP_{Cr}-4}{6}} \qquad (3)$$

where $QP_Y$, $QP_{Cb}$ and $QP_{Cr}$ correspond to the integer luma and chroma Cb and Cr QP values, respectively; they are defined in (4) to (6), respectively.

**Table 1:** The first six values of *QP*, *QStep*, *m* and *s*.

| QP | 0 | 1 | 2 | 3 | 4 | 5 |
|---|---|---|---|---|---|---|
| QStep | 0.6300 | 0.7071 | 0.7937 | 0.8909 | 1.0000 | 1.1225 |
| m | 26214 | 23302 | 20560 | 18396 | 16384 | 14564 |
| s | 40 | 45 | 51 | 57 | 64 | 72 |

$$QP_Y(QStep_Y) = \left[6 \times \log_2(QStep_Y)\right] + 4 \qquad (4)$$

$$QP_{Cb}(QStep_{Cb}) = \left[6 \times \log_2(QStep_{Cb})\right] + 4 \qquad (5)$$

$$QP_{Cr}(QStep_{Cr}) = \left[6 \times \log_2(QStep_{Cr})\right] + 4 \qquad (6)$$

In terms of the quantisation of luma and chroma transform coefficients in HEVC and the association of the QP and QStep with the Multiplication Factor (MF) and the Scaling Factor (SF), the quantised transform coefficient within an $N \times N$ TB, denoted as *t*, is computed in (7):

$$t = \frac{C \cdot (m + o)}{2^{21 + \frac{QP}{6} - \log_2 N}} \qquad (7)$$

where *C* denotes the transform coefficient, *m* corresponds to the MF associated with the QP and *o* refers to the offset corresponding to the error level incurred by quantisation rounding including the level of deadzone; $o = 2^{18}$. Variable *N* denotes the *N* value of an $N \times N$ TB [1, 2]. QStep values are integer approximated in HEVC. The inverse quantised transform coefficient, denoted as *C'*, is computed in (8):

$$C' = \frac{t \cdot s \cdot 2^{\frac{QP}{6}}}{2^{\log_2 N - 1}} \qquad (8)$$

where *s* is the SF employed for inverse quantization. The URQ method in HEVC is designed such that coefficients in a TB are equally quantised according to the frame level QP; therefore, a single QP value is applied to an entire TB of transform coefficients. MF *m* and SF *s* can be computed in (9) and (10), respectively.

$$m \approx \left[\frac{2^{14}}{QStep}\right] \qquad (9)$$

$$s = \left[2^6 \times QStep\right] \qquad (10)$$

Due to the MF and the associated bitwise operations, the values associated with quantisation and inverse quantisation are quantified without the need for divisions and floating point operations. Moreover, as shown in Table 1, the MF is inversely proportional to the QP and the QStep. Therefore, any decreases to the MF will induce greater levels of quantisation. One main objective is to ensure that an increment of *QP* (i.e., *QP* + 1) equates to an increase of *QStep* by approximately 12%.

RDOQ, which is dependent on URQ, is enabled by default in the JCT-VC HEVC HM software [11, 12]; it is, therefore, the default quantisation technique when following the common test conditions. RDOQ is a soft decision quantisation method that individually quantises coefficients in both luma and chroma TBs. This is achieved by minimising the rate-distortion Lagrangian cost function [13]. RDOQ is designed to search for an optimal set of quantised coefficients in order to establish a suitable trade-off between bitrate and quantised induced distortion; as such, a calculation for each transform coefficient is performed separately. In essence, RDOQ manipulates the quantised transform coefficients according to the final RD performance [6, 7]; therefore, it significantly outperforms URQ in terms of reducing bitrates.

In luma and chroma TBs of size *N×N*, each transform coefficient *C* with RDOQ is quantised to three level values, which are as follows: 0, $l_1$ and $l_2$. According to [6], for each transform coefficient position in a TB, the Lagrangian cost of each value of *C* is calculated when the quantisation level value, denoted as $l_i$, is equal to 0, $l_1$ or $l_2$. When *C* is quantised to value $l_i$, the Lagrangian cost $J(\lambda, l_i)$ is computed as follows in (11):

$$J(\lambda, l_i) = r(C, l_i) + \lambda \cdot b(l_i) \tag{11}$$

where $\lambda$ denotes the Lagrangian multiplier (the value is computed in [6], where *r* denotes the quantisation error if *C* is quantised to level $l_i$ and where *b* corresponds to the number of bits required to code $l_i$. Variables $l_1$ and $l_2$ are computed in (12) and (13), respectively.

$$l_1 = \left\lfloor |C| \cdot \frac{m}{2^{15+\frac{QP}{6}}} \right\rfloor \tag{12}$$

$$l_2 = l_1 + 1 \tag{13}$$

Recall that *m* is the MF, as computed in (9). The final quantised level, denoted as *q*, is computed in (14). Therefore, the Lagrangian cost function is updated to $J(\lambda, q)$.

$$q = \arg\min J(\lambda, l_i) \tag{14}$$

| DC<br>$d = 0.0000$<br>$w = 1.0000$ | AC<br>$d = 0.2357$<br>$w = 0.9460$ | AC<br>$d = 0.4714$<br>$w = 0.8007$ | AC<br>$d = 0.7071$<br>$w = 0.6065$ |
|---|---|---|---|
| AC<br>$d = 0.2357$<br>$w = 0.9460$ | AC<br>$d = 0.3333$<br>$w = 0.8948$ | AC<br>$d = 0.5271$<br>$w = 0.7575$ | AC<br>$d = 0.7454$<br>$w = 0.5737$ |
| AC<br>$d = 0.4714$<br>$w = 0.8007$ | AC<br>$d = 0.5271$<br>$w = 0.7575$ | AC<br>$d = 0.6667$<br>$w = 0.6412$ | AC<br>$d = 0.8498$<br>$w = 0.4857$ |
| AC<br>$d = 0.7071$<br>$w = 0.6065$ | AC<br>$d = 0.7454$<br>$w = 0.5737$ | AC<br>$d = 0.8498$<br>$w = 0.4857$ | AC<br>$d = 1.0000$<br>$w = 0.3679$ |

**Figure 1:** A graphical representation of the normalised Euclidean distance $d$ and weight $w$ associated with transform coefficient locations in a 4×4 TB. The darker orange, lighter orange, yellow and grey squares correspond to the DC coefficient, low frequency AC coefficients, medium AC frequency coefficients and high frequency AC coefficients, respectively.

A notable drawback of RDOQ is the computational complexity associated with the rate-distortion decisions it carries out. Recall that it is designed primarily to select an optimal quantisation level, with (11), to find a suitable trade-off between rate and distortion; this process alone requires significant computational complexity [7].

**3.0 Proposed FDPQ Technique**

FDPQ has been designed to achieve visually lossless coding at low bitrates. By decreasing the MF, the corresponding larger QStep value is employed primarily for quantising high frequency AC coefficients more aggressively, thus giving rise to perceptual quantisation. More specifically, the level of quantisation applied to each luma and chroma transform coefficient is modified indirectly by weighing — and subsequently decreasing — the corresponding MF value according to a normalised Euclidean distance parameter (see an example in Figure 1) [8]. The distance parameter is a constituent of an exponentially decreasing function. That is, it is employed for decreasing the MF and subsequently increasing the levels of quantisation applied to each AC coefficient.

As previously mentioned, FDPQ quantises high frequency AC coefficients much more aggressively than the corresponding DC coefficient and the low frequency AC coefficients. The level of quantisation applied to an individual AC coefficient is modified according to its position in relation to its Euclidean distance from the DC coefficient. An important feature of FDPQ is the fact that the distance parameter values change according to the size of the luma and chroma TBs (from 4×4 to 32×32 samples per TB), which constitutes an adaptive component of the proposed method. In terms of experimentation: due to the ubiquitous adoption of RDOQ, FDPQ is compared with RDOQ in the objective and subjective evaluations. Moreover, comparing the performance of FDPQ with RDOQ also implies comparing FDPQ with URQ; this is because RDOQ always significantly outperforms URQ.

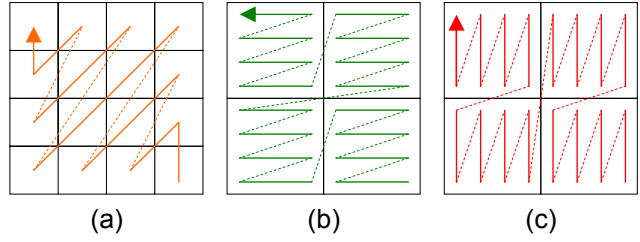

(a)               (b)              (c)

**Figure 2:** The transform coefficient scan patterns employed in HEVC. Subfigure (a) shows a diagonal reverse scan pattern used to process coefficients within a 4×4 TB. Subfigures (b) and (c) show the horizontal and vertical reverse scan patterns, respectively, for the processing of the constituent 4×4 SBs in 8×8 TBs.

The proposed technique reduces non-zero quantised coefficients. Therefore, a significant decrease in bitrates can be accomplished without incurring a perceptually discernible reduction in reconstruction quality; this is confirmed in the objective and subjective evaluations. To reiterate, FDPQ is designed with the core objective of modifying the MF $m$, such that the resulting modified MF value, denoted as $m'$, increases the QStep applied to each AC transform coefficient according to its distance from the DC coefficient, thus allowing for an efficient encoder side implementation. Note that, in the proposed FDPQ technique, modified MF $m'$ in (15) replaces unmodified MF $m$ in (9). The modified MF $m'$ is computed in (15):

$$m' = \left\lfloor \frac{2^{14}}{QStep} \right\rfloor \cdot w \qquad (15)$$

where $w$ corresponds to an exponentially decreasing function weight. Recall that $w$ modifies the MF for transform coefficients in both luma and chroma TBs; weight $w$ is quantified in (16):

$$w(d) = e^{-(d)^2} \in [0,1] \qquad (16)$$

where $d$ is the normalised Euclidean distance between the position of the current AC transform coefficient in an $N \times N$ TB. Variable $d$ is calculated in (17):

$$d = \sqrt{\frac{(x_1 - x_2)^2 + (y_1 - y_2)^2}{(x_1 - x_{max})^2 + (y_1 - y_{max})^2}} \in [0,1] \qquad (17)$$

where $(x_1, y_1)$, $(x_2, y_2)$ and $(x_{max}, y_{max})$ represent the $(x,y)$ coordinates of the DC coefficient, the current coefficient and the farthest AC coefficient, respectively. The DC coefficient is at position $x = 0$, $y = 0$ in both luma and chroma TBs.

The proposed technique is suitable for, and can be utilised with, the current scan patterns used for CABAC entropy coding in HEVC (see Figure 2). Furthermore, it is important to reiterate the fact that FDPQ is compatible for utilisation with luma and chroma TBs of any size (i.e., from 4×4 samples to 32×32 samples). The SF for the inverse quantisation process, denoted as $s'$, is computed in (18):

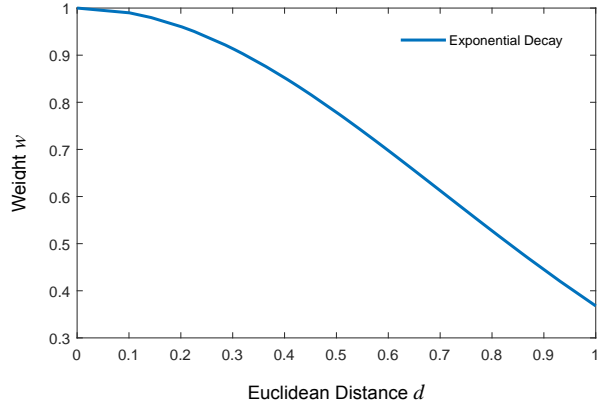

**Figure 3:** A plot showing the exponential decay in relation to weight *w* with respect to Euclidean distance *d*. Note *w* decreases as the distance increases. In addition, the curve corresponds very closely to the MTF of the HVS.

$$s' = \left(\frac{2^{20}}{m}\right) \cdot e^{-(d)^2} \qquad (18)$$

Note that the values of *m* in (9) and *w* in (16) are available at the decoder side. MF *m* is known from the bitstream after entropy decoding with CABAC; moreover, distance parameter *d* is determined by the transform coefficient location in luma and chroma TBs. Therefore, the value for *s'* employed in the encoder loop for generating reference frames is the same value as *s'* at the decoder side. As such, this allows for an efficient and low complexity encoder side implementation of FDPQ; i.e., all of the values associated with FDPQ are signalled to the decoder in the Picture Parameter Set (PPS).

In relation to the FDPQ, note that the quantisation and inverse quantisation procedures are identical for all TB sizes (i.e., from 4×4 samples to 32×32 samples). It is important to mention that weight *w*, by decreasing the MF, indirectly increases the QP value for each AC transform coefficient without the need to analyse multiple QPs (as is the case with RDOQ). Consequently, this can give rise to improvements in terms of computational complexity reductions and runtime decreases.

In addition to the primary objective of achieving perceptual quantisation (i.e., significant decreases in bitrates compared with RDOQ and URQ), another objective is to ensure that computational complexity and the associated encoding and decoding runtimes are not increased. As is the case with URQ, the computational complexity of FDPQ is computed in linear time *T*, as computed in (19):

$$T(n) = O(n) \qquad (19)$$

Like URQ, the computational performance of FDPQ is directly proportional to the number of transform coefficients being processed in each luma and chroma TB. FDPQ reduces the number of non-zero quantised transform coefficients, which may give rise to faster entropy coding and decoding times compared with RDOQ. This is because fewer non-zero coefficients results in an encoded bitstream with fewer bits.

## 4.0 Experimental Evaluations, Results and Discussion

It is important to note that the following evaluation procedure was previously utilised in the unpublished thesis in [9] and, in addition, in our previously published work in [14]. With this mind, FDPQ is evaluated and compared with RDOQ [6, 7], which, to reiterate, is the default quantisation technique employed in HEVC. Note that RDOQ, by definition, always significantly outperforms URQ in terms of coding efficiency. Therefore, comparing FDPQ with RDOQ also implies comparing FDPQ with URQ. To evaluate the efficacy of FDPQ, an exhaustive evaluation procedure is undertaken. The objective visual quality evaluations correspond, as closely as possible, to the Common Test Conditions and Software Reference Configurations recommended by JCT-VC [15, 16]. The experimental setup utilised in this paper includes testing the proposed technique with five initial QP data points (i.e., QPs 17, 22, 27, 32 and 37) and the All Intra (AI) and Random Access (RA) encoding configurations [15, 16].

Because the proposed technique is optimised for perceptual compression, it is of significant importance to undertake exhaustive subjective visual quality evaluations (in addition to the aforementioned objective visual quality evaluations). Therefore, a United Nations' ITU-R standardised subjective evaluation procedure [17] is employed. The subjective visual quality evaluations are the most important set of experiments in terms of measuring the visual quality of perceptually compressed video sequences (especially so when targeting low bitrates in visually lossless coding techniques).

In the objective visual quality evaluations, the Structural Similarity Index (SSIM) [18] and the PSNR visual quality metrics are employed to assess the reconstruction quality of the compressed video frames; i.e., in the sequences coded by FDPQ and the reference technique (RDOQ). The SSIM and PSNR values presented in this paper correspond to objective spatial reconstruction analyses of the intra-frames (in the AI tests) and also the inter-frames — in addition to the intra key frames — (in the RA tests), per sequence. In other words, the mean SSIM and PSNR values are computed by comparing the coded frames, using initial QPs 17, 22, 27, 32 and 37, with the frames in the raw data; this is carried out in bitmap image form for each sequence and each QP.

FDPQ is implemented into the JCT-VC HEVC HM 16.7 reference software [11, 12]; it is evaluated on 18 official test sequences provided by JCT-VC. The test sequences are as follows: the YCbCr 4:2:0, 4:2:2 and 4:4:4 versions of the BirdsInCage, DuckAndLegs, Kimono, OldTownCross, ParkScene and Traffic sequences. Note that Figures 4 - 9 show a frame from each 4:4:4 raw sequence in bitmap image form. All of the aforementioned sequences comprise a spatial resolution of full High Definition (HD), 1920×1080 pixels (1080p). The 4:4:4 and 4:2:2 versions of these sequences contain a higher dynamic range (i.e., 10-bits per sample per channel, which equates to 30-bits per sample), whereas the 4:2:0 versions comprise 8-bits per sample per channel (24-bits per sample).

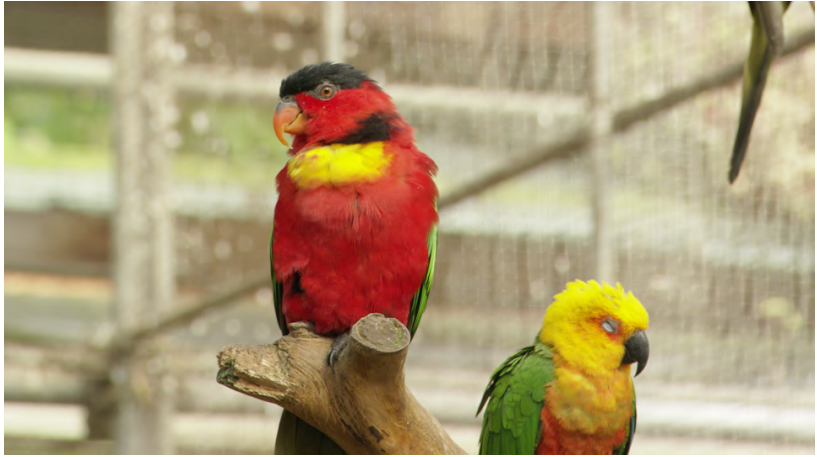

**Figure 4:** JCT-VC BirdsInCage 4:4:4 Raw Data (HD 1080p) for HEVC Evaluations

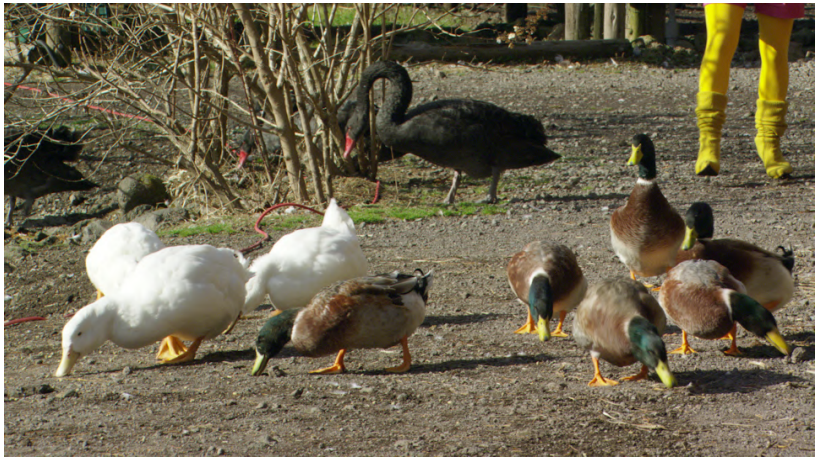

**Figure 5:** JCT-VC DuckAndLegs 4:4:4 Raw Data (HD 1080p) for HEVC Evaluations

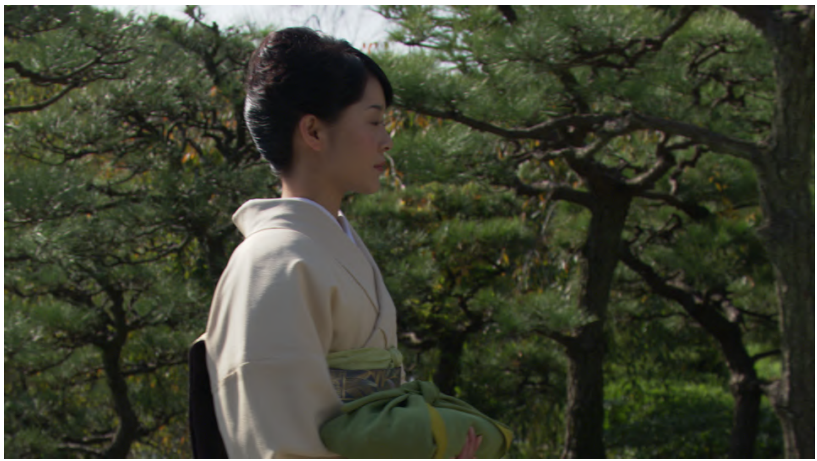

**Figure 6:** JCT-VC Kimono 4:4:4 Raw Data (HD 1080p) for HEVC Evaluations

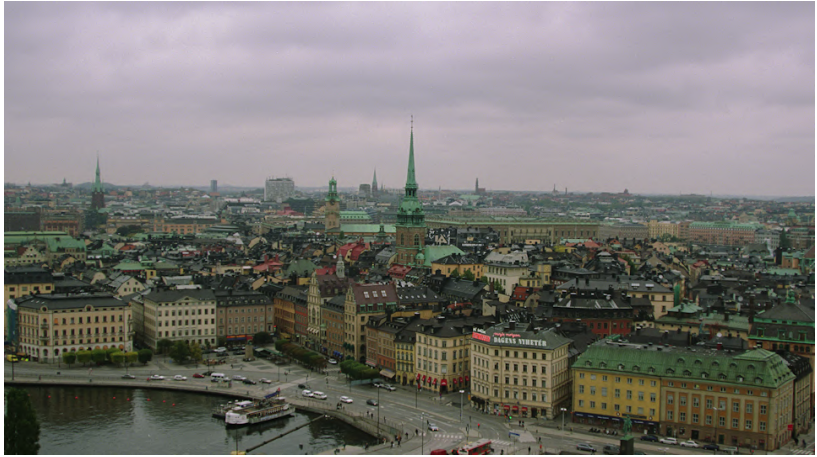

**Figure 7:** JCT-VC OldTownCross 4:4:4 Raw Data (HD 1080p) for HEVC Evaluations

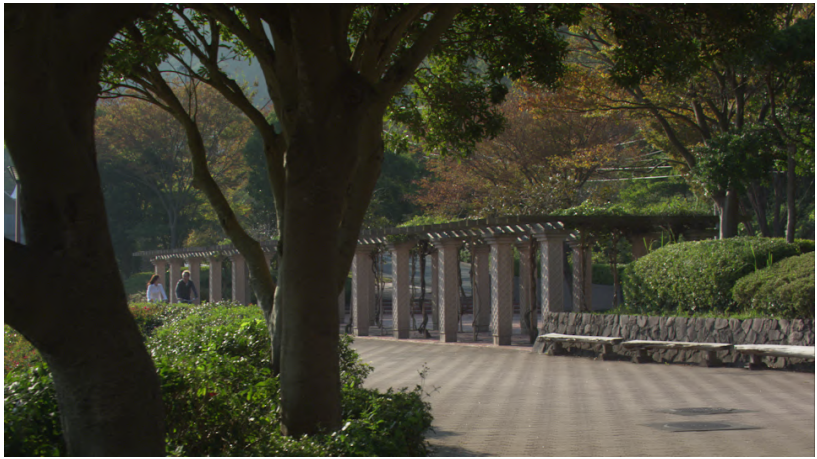

**Figure 8:** JCT-VC ParkScene 4:4:4 Raw Data (HD 1080p) for HEVC Evaluations

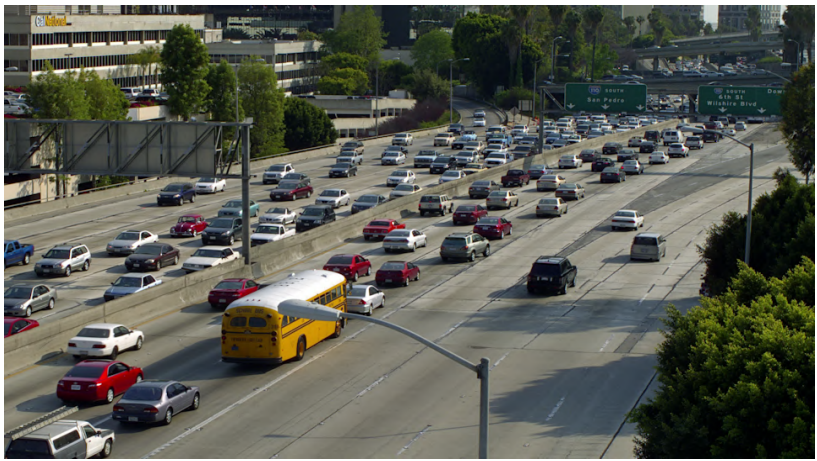

**Figure 9:** JCT-VC Traffic 4:4:4 Raw Data (HD 1080p) for HEVC Evaluations

**Table 2:** The criteria for quantifying the MOS with respect to the visual reconstruction quality of a compressed video sequence (compared with the raw video data).

| MOS | Visual Quality Difference |
|---|---|
| 5 | Imperceptible (Visually Lossless) |
| 4 | Very Slightly Perceptible |
| 3 | Moderately Perceptible |
| 2 | Significantly Perceptible |
| 1 | Extremely Obvious |

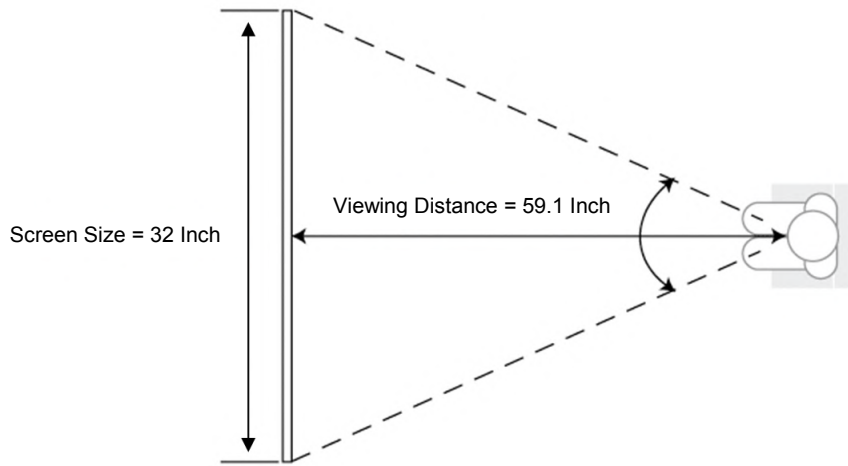

**Figure 10:** The viewing conditions employed in the subjective evaluations. This figure illustrates the screen size of the display and the viewing distance of the participant from the display.

The desktop PC hardware utilised in the experimental setup consists of the following:

- Intel Core i7-4770 CPU (3.4 GHz per core);
- DDR3 SDRAM (24 GB, 680 MHz);
- NVIDIA GeForce 750 Ti (2 GB DDR5 SDRAM, Ultra HD 4K and HDR Capable);
- TV/Visual Display Unit (VDU): HD 1080p 32 " Samsung F5500 LED Smart TV.

On the aforementioned hardware used in the experimental setup, note that, due to the higher dynamic ranges and the increased levels of chroma saturation in the raw YCbCr 4:4:4 and 4:2:2 10-bit sequences, the superior visual quality of this data is perceptually discernible in comparison with the chroma subsampled raw 4:2:0 8-bit sequences.

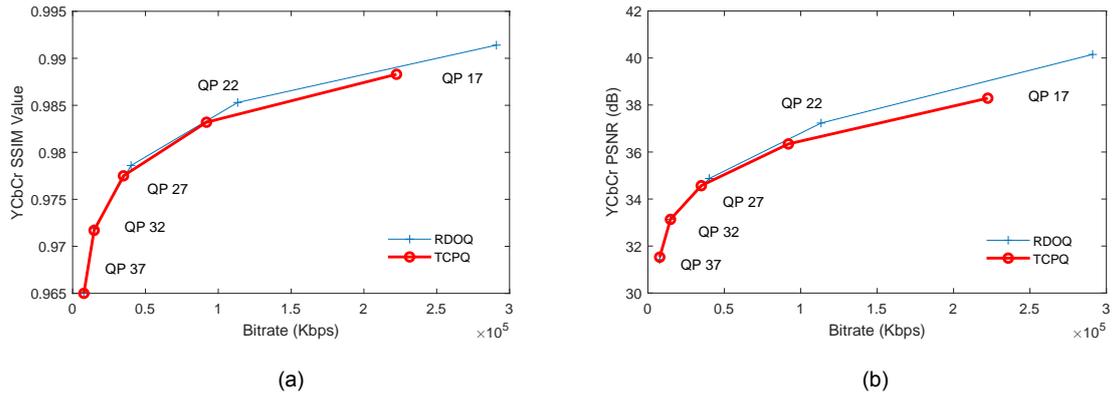

**Figure 11:** Two plots which highlight the bitrate reductions attained by FDPQ compared with RDOQ. The subfigures show the bitrate reductions achieved by FDPQ on the following sequences. Subfigure (a): BirdsInCage 4:4:4 (AI - SSIM). Subfigure (b): BirdsInCage 4:4:4 (AI - PSNR).

The main objectives of the subjective evaluations are as follows: i) to establish if the proposed technique achieve visually lossless coding — compared with the raw video data — at low bitrates; ii) to ascertain if the proposed technique outperforms, or matches, the performance of the reference technique (RDOQ) in terms of visual quality. To reiterate, the directions of the internationally standardised United Nations' ITU subjective evaluations entitled: *Subjective Video Quality Assessment Methods* (ITU-R Rec. P.910 [17]) are followed as closely as possible. In ITU-R Rec. P.910, the following conditions are recommended: Number of Individuals $\geq 4$ and $\leq 40$; Viewing Distance: $1\text{-}8 \times H$, where $H$ is the height of the TV/VDU; Quantify the Mean Opinion Score (MOS) — see Table 2; Spatial and Temporal Video Analysis.

In accordance with the recommended conditions specified in the ITU-R Rec. P.910 subjective evaluation procedures, four individuals participated in an exhaustive assessment (AI and RA tests at QP 22, 27, 32 and 37). Note that two additional individuals participated in preliminary subjective evaluations (AI and RA tests at QPs 17 and 22 only). The viewing distance of the participants from the TV/VDU is 1.5m in all evaluations (1.5m ≈ 59.1 inch). The height $H$ of the TV/VDU is 15.7 inch and the viewing distance is approximately $4 \times H$ (see Figure 10). Concerning the two additional subjective evaluations, the objective is to ascertain if FDPQ achieves visually lossless coding (MOS = 5) in the AI and RA tests at QPs 17 and 22 only. To this end, the participants confirmed that an MOS = 5 was recorded in these additional subjective evaluations.

In terms of the bitrate reductions achieved by FDPQ versus RDOQ, Table 3 shows that the highest bitrate reductions in the AI tests are attained on the BirdsInCage 4:4:4 and OldTownCross 4:4:4 sequences — see Figure 11. Similarly, the highest overall bitrate reductions accomplished (RA tests) by FDPQ are gained on the BirdsInCage 4:4:4 and OldTownCross 4:4:4 sequences with overall bitrate decrease of 41% and 34.7%, respectively. Over five QP data points (i.e., initial QPs 17, 22, 27, 32 and 37) using the AI and RA encoding configurations, high bitrate reductions are attained by FDPQ in addition to accomplishing perceptually identical reconstruction quality.

**Table 3:** Tabulated bitrate reduction percentages attained by the proposed FDPQ technique compared with RDOQ. The bitrate reductions are averaged over five QP data points (initial QPs 17, 22, 27, 37 and 37). The AI results are shown on the left; the RA results are shown on the right. The text in red indicates bitrate inflations.

| FDPQ Versus RDOQ (YCbCr 4:2:0) – AI | | FDPQ Versus RDOQ (YCbCr 4:2:0) – RA | |
|---|---|---|---|
| **Sequence** | Bitrate (%) | **Sequence** | Bitrate (%) |
| BirdsInCage | −8.6 | BirdsInCage | −30.1 |
| DuckAndLegs | −10.8 | DuckAndLegs | −25.3 |
| Kimono | −4.7 | Kimono | −8.2 |
| OldTownCross | −14.0 | OldTownCross | −30.0 |
| ParkScene | −6.6 | ParkScene | −9.9 |
| Traffic | −3.9 | Traffic | −9.3 |
| **FDPQ Versus RDOQ (YCbCr 4:2:2) – AI** | | **FDPQ Versus RDOQ (YCbCr 4:2:2) – RA** | |
| **Sequence** | Bitrate (%) | **Sequence** | Bitrate (%) |
| BirdsInCage | −11.2 | BirdsInCage | −34.2 |
| DuckAndLegs | −13.9 | DuckAndLegs | −30.5 |
| Kimono | −10.3 | Kimono | −16.4 |
| OldTownCross | −16.7 | OldTownCross | −33.0 |
| ParkScene | −17.2 | ParkScene | −22.9 |
| Traffic | −4.0 | Traffic | −11.2 |
| **FDPQ Versus RDOQ (YCbCr 4:4:4) – AI** | | **FDPQ Versus RDOQ (YCbCr 4:4:4) – RA** | |
| **Sequence** | Bitrate (%) | **Sequence** | Bitrate (%) |
| BirdsInCage | −20.4 | BirdsInCage | −41.0 |
| DuckAndLegs | −17.0 | DuckAndLegs | −32.4 |
| Kimono | −15.4 | Kimono | −23.7 |
| OldTownCross | −19.7 | OldTownCross | −34.7 |
| ParkScene | −13.5 | ParkScene | −26.2 |
| Traffic | −4.7 | Traffic | −14.3 |

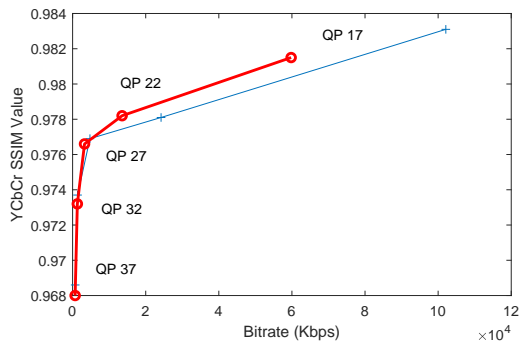
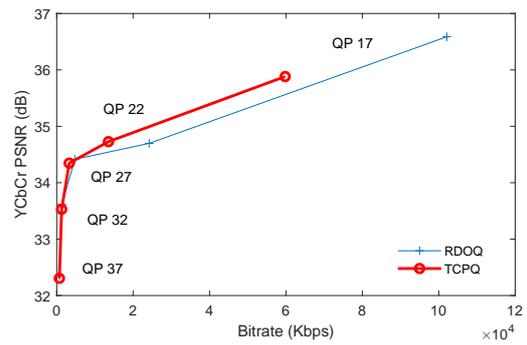

(a)      (b)

**Figure 12:** Two plots which highlight the bitrate reductions attained by FDPQ compared with RDOQ. The subfigures show the bitrate reductions achieved by FDPQ on the following sequences. Subfigure (a): BirdsInCage 4:4:4 (RA - SSIM). Subfigure (b): BirdsInCage 4:4:4 (RA - PSNR).

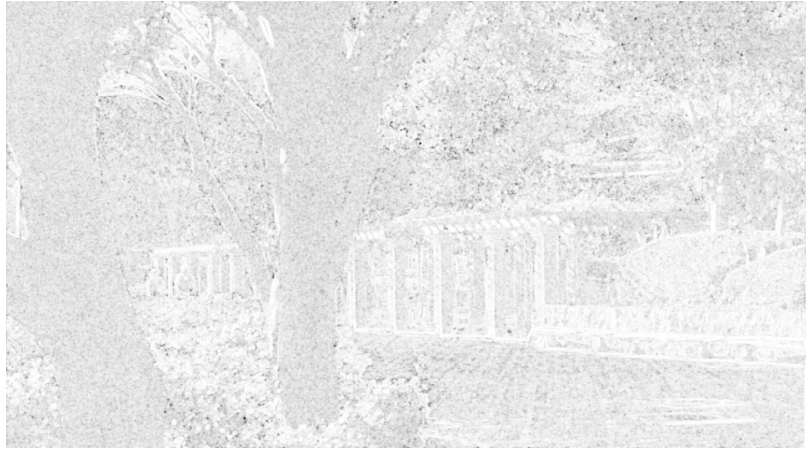

(a) Luma Channel

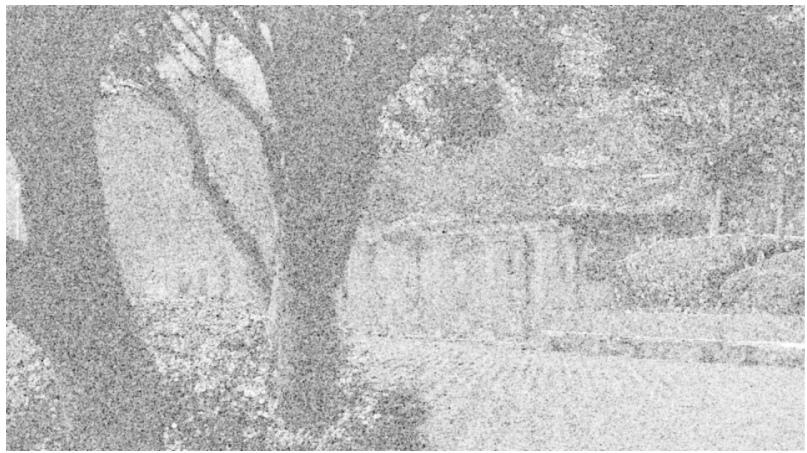

(b) Chroma Cb Channel

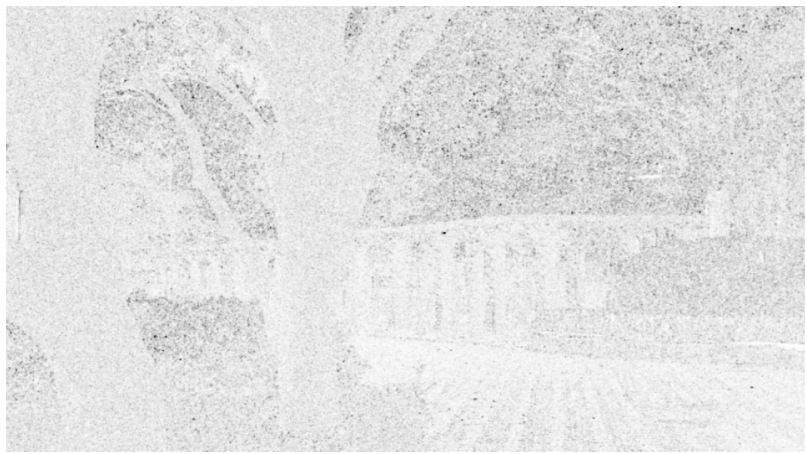

(c) Chroma Cr Channel

**Figure 13:** The SSIM Index Map (structural reconstruction errors) of a FDPQ-coded inter-frame (RA QP = 22 test) versus the raw data (ParkScene 4:4:4 sequence). In subfigures (a), (b) and (c), respectively, the luma (Y), chroma Cb and chroma Cr structural reconstruction errors are shown separately. Note that these reconstruction errors in the FDPQ-coded compressed sequence are imperceptible to the HVS according to the subjective evaluations (compare the subfigures in Figure 14).

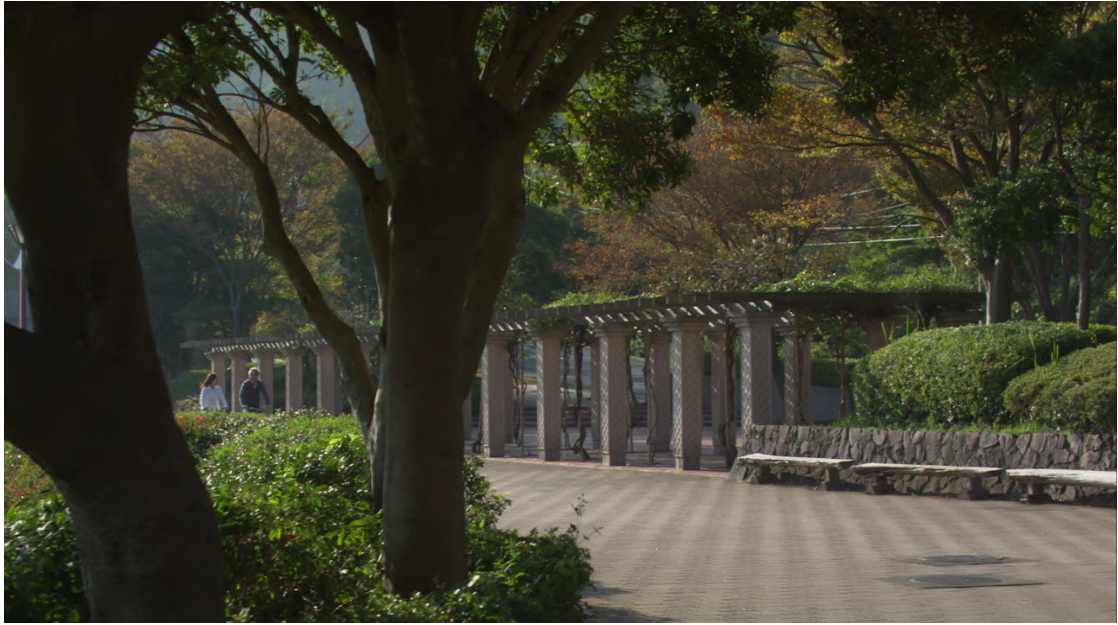

(a) FDPQ-Coded Inter-Frame (RA = QP 22): YCbCr PSNR = 32.8848 and YCbCr SSIM = 0.8629

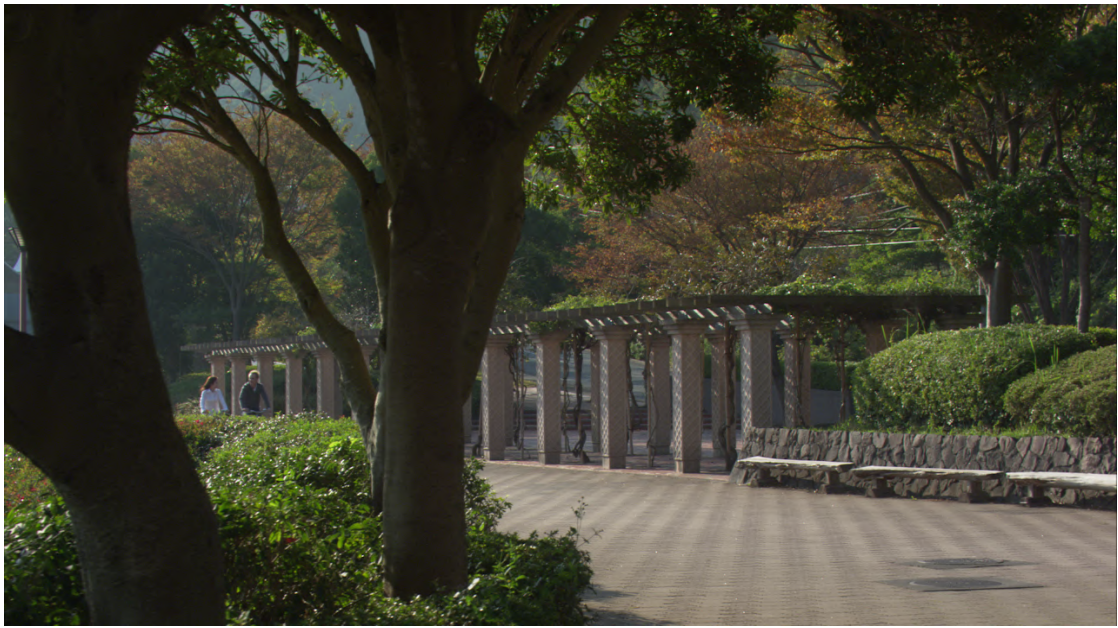

(b) Raw Data

**Figure 14:** A frame from the ParkScene 4:4:4 sequence. Subfigure (a) is a FDPQ-coded inter-frame from this sequence (RA QP = 22 test). Subfigure (b) is the corresponding raw data. Note that the FDPQ-coded sequence in (a) is perceptually indistinguishable from the raw data in (b) according to the subjective evaluations.

**Table 4:** The SSIM results for the 'FDPQ versus the raw data' in comparison with 'RDOQ versus the raw data' (initial QPs 17, 22, 27, 32 and 37) using the AI encoding configuration. Green text indicates superior results.

**Mean SSIM Values (Per Sequence, Per QP): FDPQ Versus RDOQ (YCbCr 4:2:0) – All Intra**

| Sequence | FDPQ | | | | | RDOQ | | | | |
|---|---|---|---|---|---|---|---|---|---|---|
| | QP 17 | QP 22 | QP 27 | QP 32 | QP 37 | QP 17 | QP 22 | QP 27 | QP 32 | QP 37 |
| BirdsInCage | 0.9925 | 0.9898 | 0.9863 | 0.9814 | 0.9760 | **0.9932** | **0.9902** | **0.9865** | **0.9818** | **0.9761** |
| DuckAndLegs | 0.9721 | 0.9480 | 0.9153 | 0.8826 | 0.8369 | **0.9861** | **0.9611** | **0.9187** | **0.8856** | **0.8393** |
| Kimono | 0.9569 | 0.9376 | 0.9187 | 0.8946 | 0.8644 | **0.9619** | **0.9378** | **0.9189** | **0.8962** | **0.8664** |
| OldTownCross | 0.9442 | 0.9025 | 0.8601 | 0.8290 | **0.7863** | **0.9664** | **0.9152** | **0.8627** | **0.8297** | 0.7856 |
| ParkScene | 0.9683 | 0.9509 | 0.9255 | 0.8873 | 0.8379 | **0.9728** | **0.9557** | **0.9292** | **0.8901** | **0.8375** |
| Traffic | 0.9769 | 0.9632 | 0.9395 | 0.9045 | 0.8551 | **0.9799** | **0.9656** | **0.9418** | **0.9056** | **0.8542** |

**Mean SSIM Values (Per Sequence, Per QP): FDPQ Versus RDOQ (YCbCr 4:2:2) – All Intra**

| Sequence | FDPQ | | | | | RDOQ | | | | |
|---|---|---|---|---|---|---|---|---|---|---|
| | QP 17 | QP 22 | QP 27 | QP 32 | QP 37 | QP 17 | QP 22 | QP 27 | QP 32 | QP 37 |
| BirdsInCage | 0.9913 | 0.9881 | 0.9843 | 0.9787 | 0.9709 | **0.9926** | **0.9888** | **0.9844** | **0.9788** | **0.9713** |
| DuckAndLegs | 0.9698 | 0.9422 | 0.8985 | 0.8542 | 0.8014 | **0.9877** | **0.9644** | **0.9066** | **0.8564** | **0.8028** |
| Kimono | 0.9487 | 0.9213 | 0.8981 | 0.8711 | 0.8310 | **0.9615** | **0.9236** | **0.8981** | **0.8718** | **0.8319** |
| OldTownCross | 0.9386 | 0.8838 | 0.8299 | 0.7955 | **0.7480** | **0.9682** | **0.9053** | **0.8300** | **0.7970** | 0.7474 |
| ParkScene | 0.9407 | 0.9015 | 0.8635 | 0.8257 | 0.7774 | **0.9626** | **0.9129** | **0.8654** | **0.8265** | **0.7775** |
| Traffic | 0.9786 | 0.9650 | 0.9420 | 0.9032 | **0.8407** | **0.9816** | **0.9673** | **0.9437** | **0.9044** | 0.8391 |

**Mean SSIM Values (Per Sequence, Per QP): FDPQ Versus RDOQ (YCbCr 4:4:4) – All Intra**

| Sequence | FDPQ | | | | | RDOQ | | | | |
|---|---|---|---|---|---|---|---|---|---|---|
| | QP 17 | QP 22 | QP 27 | QP 32 | QP 37 | QP 17 | QP 22 | QP 27 | QP 32 | QP 37 |
| BirdsInCage | 0.9883 | 0.9832 | 0.9775 | 0.9717 | 0.9650 | **0.9914** | **0.9853** | **0.9786** | **0.9719** | **0.9651** |
| DuckAndLegs | 0.9718 | 0.9473 | 0.8949 | 0.8301 | 0.7751 | **0.9900** | **0.9718** | **0.9201** | **0.8338** | **0.7771** |
| Kimono | 0.9494 | 0.9145 | **0.8793** | 0.8512 | 0.8146 | **0.9653** | **0.9245** | 0.8786 | **0.8533** | **0.8161** |
| OldTownCross | 0.9425 | 0.8810 | 0.7882 | 0.7318 | **0.6861** | **0.9745** | **0.9198** | **0.7978** | **0.7330** | 0.6860 |
| ParkScene | 0.9469 | 0.9047 | 0.8547 | 0.8134 | 0.7663 | **0.9669** | **0.9205** | **0.8571** | **0.8146** | **0.7669** |
| Traffic | 0.9798 | 0.9657 | 0.9425 | 0.9063 | 0.8483 | **0.9828** | **0.9684** | **0.9452** | **0.9082** | **0.8485** |

The subfigures in Figure 13 highlight the structural luma and chroma reconstruction errors in the 'FDPQ versus the raw data' test conducted on the ParkScene 4:4:4 sequence. In comparison with the raw data, the structural reconstruction errors incurred by FDPQ in the Y, Cb and Cr channels are concentrated mostly in the high variance regions in each channel. In spite of these quantisation-induced errors in the FDPQ-coded sequence, visually lossless coding is achieved in the AI and RA QP = 22 tests, as confirmed in the subjective evaluations. Compare the subfigures in Figure 14 for a visual example.

**Table 5:** The SSIM results for the 'FDPQ versus the raw data' in comparison with 'RDOQ versus the raw data' (initial QPs 17, 22, 27, 32 and 37) using the RA encoding configuration. Green text indicates superior results.

**Mean SSIM Values (Per Sequence, Per QP): FDPQ Versus RDOQ (YCbCr 4:2:0) – Random Access**

| Sequence | FDPQ | | | | | RDOQ | | | | |
|---|---|---|---|---|---|---|---|---|---|---|
| | QP 17 | QP 22 | QP 27 | QP 32 | QP 37 | QP 17 | QP 22 | QP 27 | QP 32 | QP 37 |
| BirdsInCage | 0.9904 | 0.9894 | 0.9872 | 0.9836 | 0.9797 | **0.9905** | **0.9897** | **0.9877** | **0.9843** | **0.9801** |
| DuckAndLegs | 0.9371 | 0.9173 | 0.9001 | 0.8734 | 0.8320 | **0.9415** | **0.9208** | **0.9055** | **0.8791** | **0.8384** |
| Kimono | 0.9320 | 0.9236 | 0.9087 | 0.8874 | 0.8640 | **0.9324** | **0.9253** | **0.9111** | **0.8912** | **0.8668** |
| OldTownCross | 0.8856 | 0.8641 | 0.8531 | 0.8347 | 0.8001 | **0.8890** | **0.8648** | **0.8559** | **0.8386** | **0.8047** |
| ParkScene | 0.9587 | 0.9450 | 0.9213 | 0.8859 | 0.8396 | **0.9629** | **0.9505** | **0.9278** | **0.8936** | **0.8460** |
| Traffic | 0.9719 | 0.9619 | 0.9429 | 0.9132 | 0.8695 | **0.9746** | **0.9652** | **0.9468** | **0.9174** | **0.8745** |

**Mean SSIM Values (Per Sequence, Per QP): FDPQ Versus RDOQ (YCbCr 4:2:2) – Random Access**

| Sequence | FDPQ | | | | | RDOQ | | | | |
|---|---|---|---|---|---|---|---|---|---|---|
| | QP 17 | QP 22 | QP 27 | QP 32 | QP 37 | QP 17 | QP 22 | QP 27 | QP 32 | QP 37 |
| BirdsInCage | 0.9880 | **0.9867** | 0.9848 | 0.9811 | 0.9755 | **0.9882** | **0.9867** | **0.9850** | **0.9813** | **0.9759** |
| DuckAndLegs | 0.9324 | 0.8939 | 0.8763 | 0.8465 | 0.8002 | **0.9463** | **0.8965** | **0.8819** | **0.8532** | **0.8074** |
| Kimono | 0.9085 | **0.9001** | 0.8873 | 0.8667 | 0.8372 | **0.9072** | **0.9001** | **0.8891** | **0.8693** | **0.8408** |
| OldTownCross | 0.8532 | **0.8222** | 0.8136 | 0.7955 | 0.7567 | **0.8541** | 0.8221 | **0.8161** | **0.7987** | **0.7606** |
| ParkScene | 0.8934 | 0.8797 | 0.8603 | 0.8299 | 0.7863 | **0.8972** | **0.8835** | **0.8653** | **0.8339** | **0.7908** |
| Traffic | 0.9713 | 0.9614 | 0.9436 | 0.9122 | 0.8585 | **0.9740** | **0.9648** | **0.9476** | **0.9165** | **0.8660** |

**Mean SSIM Values (Per Sequence, Per QP): FDPQ Versus RDOQ (YCbCr 4:4:4) – Random Access**

| Sequence | FDPQ | | | | | RDOQ | | | | |
|---|---|---|---|---|---|---|---|---|---|---|
| | QP 17 | QP 22 | QP 27 | QP 32 | QP 37 | QP 17 | QP 22 | QP 27 | QP 32 | QP 37 |
| BirdsInCage | 0.9815 | **0.9782** | 0.9766 | 0.9732 | 0.9680 | **0.9831** | 0.9781 | **0.9769** | **0.9737** | **0.9686** |
| DuckAndLegs | 0.9377 | 0.8853 | 0.8477 | 0.8181 | 0.7697 | **0.9610** | **0.8944** | **0.8516** | **0.8234** | **0.7769** |
| Kimono | 0.8994 | **0.8740** | 0.8629 | 0.8446 | 0.8172 | **0.9008** | 0.8736 | **0.8646** | **0.8474** | **0.8218** |
| OldTownCross | 0.8465 | **0.7512** | 0.7457 | 0.7292 | 0.6937 | **0.8715** | 0.7485 | **0.7477** | **0.7328** | **0.6986** |
| ParkScene | 0.8959 | 0.8629 | 0.8451 | 0.8167 | 0.7739 | **0.9016** | **0.8644** | **0.8492** | **0.8214** | **0.7789** |
| Traffic | 0.9697 | 0.9599 | 0.9426 | 0.9134 | 0.8641 | **0.9725** | **0.9633** | **0.9468** | **0.9179** | **0.8705** |

In terms of the mathematical reconstruction quality of the compressed video data, as shown in Table 4 and Table 5, the SSIM values for the FDPQ-coded sequences are not significantly different from those of the RDOQ-coded sequences. In most cases, the SSIM values for the RDOQ-coded sequences are higher. However, these results do not reflect how the subjective evaluation participants perceived the visual quality of the compressed video sequences. That is, the differences in mathematical reconstruction quality, as quantified by SSIM, are perceptually indiscernible according to the subjective evaluations conducted.

**Table 6:** The PSNR results for the 'FDPQ versus the raw data' in comparison with 'RDOQ versus the raw data' (initial QPs 17, 22, 27, 32 and 37) using the AI encoding configuration. Green text indicates superior results.

**Mean PSNR (dB) Per Sequence, Per QP: FDPQ Versus RDOQ (YCbCr 4:2:0) – All Intra**

| Sequence | FDPQ | | | | | RDOQ | | | | |
|---|---|---|---|---|---|---|---|---|---|---|
| | QP 17 | QP 22 | QP 27 | QP 32 | QP 37 | QP 17 | QP 22 | QP 27 | QP 32 | QP 37 |
| BirdsInCage | 40.2296 | 38.4002 | 36.6354 | 34.7860 | 32.8728 | 41.0129 | 38.7012 | 36.6945 | 34.7871 | 32.7339 |
| DuckAndLegs | 36.1747 | 33.5289 | 31.0645 | 29.0790 | 27.1250 | 39.3833 | 34.8995 | 31.4116 | 29.2380 | 27.1996 |
| Kimono | 39.1947 | 37.3183 | 35.5245 | 33.6365 | 31.6790 | 39.6659 | 37.3840 | 35.5691 | 33.7496 | 31.7134 |
| OldTownCross | 36.8165 | 34.3910 | 32.4754 | 30.9350 | 29.2072 | 39.1555 | 35.1321 | 32.6708 | 31.0068 | 29.1971 |
| ParkScene | 38.3919 | 35.9489 | 33.5807 | 31.2039 | 28.9970 | 40.2817 | 37.0864 | 34.0211 | 31.3811 | 29.0183 |
| Traffic | 39.0849 | 36.7575 | 34.1328 | 31.6338 | 29.2000 | 40.5629 | 37.5261 | 34.5096 | 31.7751 | 29.2048 |

**Mean PSNR (dB) Per Sequence, Per QP: FDPQ Versus RDOQ (YCbCr 4:2:2) – All Intra**

| Sequence | FDPQ | | | | | RDOQ | | | | |
|---|---|---|---|---|---|---|---|---|---|---|
| | QP 17 | QP 22 | QP 27 | QP 32 | QP 37 | QP 17 | QP 22 | QP 27 | QP 32 | QP 37 |
| BirdsInCage | 39.5318 | 37.7115 | 36.0440 | 34.2490 | 32.1414 | 40.6194 | 38.1548 | 36.0751 | 34.2352 | 32.0200 |
| DuckAndLegs | 35.5482 | 32.8397 | 30.3098 | 28.2637 | 26.2842 | 39.4249 | 34.8383 | 30.7462 | 28.3535 | 26.3148 |
| Kimono | 38.3566 | 36.4261 | 34.8801 | 33.0928 | 30.9228 | 39.4174 | 36.5508 | 34.8996 | 33.1495 | 30.9463 |
| OldTownCross | 36.2502 | 33.6149 | 31.7518 | 30.3369 | 28.6096 | 39.1385 | 34.5342 | 31.8470 | 30.3877 | 28.5875 |
| ParkScene | 36.8270 | 34.5044 | 32.4421 | 30.4694 | 28.5484 | 39.1704 | 35.3941 | 32.7094 | 30.5337 | 28.5243 |
| Traffic | 39.3220 | 36.9278 | 34.2833 | 31.5394 | 28.8084 | 40.9430 | 37.7508 | 34.6483 | 31.6633 | 28.7790 |

**Mean PSNR (dB) Per Sequence, Per QP: FDPQ Versus RDOQ (YCbCr 4:4:4) – All Intra**

| Sequence | FDPQ | | | | | RDOQ | | | | |
|---|---|---|---|---|---|---|---|---|---|---|
| | QP 17 | QP 22 | QP 27 | QP 32 | QP 37 | QP 17 | QP 22 | QP 27 | QP 32 | QP 37 |
| BirdsInCage | 38.2903 | 36.3439 | 34.5694 | 33.1340 | 31.5270 | 40.1517 | 37.2325 | 34.8728 | 33.1364 | 31.4187 |
| DuckAndLegs | 35.2366 | 32.6017 | 29.7395 | 27.6077 | 25.8644 | 39.6440 | 35.1159 | 30.6737 | 27.7083 | 25.9131 |
| Kimono | 38.0461 | 35.8240 | 34.1624 | 32.6565 | 30.7349 | 39.4838 | 36.2514 | 34.1785 | 32.7523 | 30.7730 |
| OldTownCross | 35.8234 | 32.8982 | 30.5346 | 29.1914 | 27.8360 | 39.3534 | 34.4861 | 30.7330 | 29.2415 | 27.8500 |
| ParkScene | 37.1236 | 34.6297 | 32.4205 | 30.5104 | 28.5994 | 39.3890 | 35.5348 | 32.6720 | 30.6046 | 28.6032 |
| Traffic | 39.5682 | 37.0596 | 34.3502 | 31.6235 | 28.8940 | 41.2753 | 37.9611 | 34.7950 | 31.8105 | 28.9053 |

The proposed FDPQ method and RDOQ operate in a comparable manner. Therefore, the structural reconstruction quality is relatively similar in both FDPQ-coded sequences and RDOQ-coded sequences. Recall that RDOQ establishes a suitable trade-off between the bitrate and compression-induced distortion; this is achieved by minimising the rate-distortion Lagrangian cost function. Consequently, higher levels of quantisation are applied to high frequency AC coefficients in luma and chroma TBs. Similarly, FDPQ, with the proposed Euclidean distance parameter approach, ensures that high frequency AC coefficients are quantised to a much higher degree than the DC coefficient and the low frequency AC coefficients.

**Table 7:** The PSNR results for the 'FDPQ versus the raw data' in comparison with 'RDOQ versus the raw data' (initial QPs 17, 22, 27, 32 and 37) using the RA encoding configuration. Green text indicates superior results.

**Mean PSNR (dB) Per Sequence, Per QP: FDPQ Versus RDOQ (YCbCr 4:2:0) – Random Access**

| Sequence | FDPQ | | | | | RDOQ | | | | |
|---|---|---|---|---|---|---|---|---|---|---|
| | QP 17 | QP 22 | QP 27 | QP 32 | QP 37 | QP 17 | QP 22 | QP 27 | QP 32 | QP 37 |
| BirdsInCage | 38.6175 | 38.0313 | 36.9060 | 35.4596 | 33.8542 | 38.7047 | 38.1957 | 37.1156 | 35.6208 | 33.9313 |
| DuckAndLegs | 32.6849 | 31.1758 | 29.8329 | 28.3225 | 26.6617 | 33.0892 | 31.5517 | 30.2455 | 28.6183 | 26.8910 |
| Kimono | 37.1279 | 36.0300 | 34.4623 | 32.6953 | 31.0570 | 37.1744 | 36.1917 | 34.6534 | 32.9357 | 31.2515 |
| OldTownCross | 33.8608 | 32.8185 | 32.1905 | 31.2190 | 29.8049 | 34.1082 | 32.8888 | 32.3525 | 31.4017 | 29.9488 |
| ParkScene | 37.1455 | 35.2638 | 33.1639 | 31.0835 | 29.1460 | 38.1238 | 36.1143 | 33.7870 | 31.5525 | 29.4531 |
| Traffic | 37.9109 | 36.1445 | 34.0528 | 31.8976 | 29.7047 | 38.6447 | 36.8452 | 34.5873 | 32.2633 | 29.9602 |

**Mean PSNR (dB) Per Sequence, Per QP: FDPQ Versus RDOQ (YCbCr 4:2:2) – Random Access**

| Sequence | FDPQ | | | | | RDOQ | | | | |
|---|---|---|---|---|---|---|---|---|---|---|
| | QP 17 | QP 22 | QP 27 | QP 32 | QP 37 | QP 17 | QP 22 | QP 27 | QP 32 | QP 37 |
| BirdsInCage | 37.6014 | 36.9339 | 36.1972 | 34.8992 | 33.1689 | 37.7717 | 36.9658 | 36.3165 | 34.9791 | 33.2516 |
| DuckAndLegs | 32.0353 | 30.1167 | 29.0295 | 27.5933 | 25.9268 | 32.8429 | 30.3252 | 29.3686 | 27.8630 | 26.1294 |
| Kimono | 35.8131 | 35.0102 | 33.7945 | 32.2135 | 30.5026 | 35.7653 | 35.0352 | 33.9246 | 32.2135 | 30.7039 |
| OldTownCross | 32.8684 | 31.8850 | 31.4376 | 30.6083 | 29.2096 | 32.9684 | 31.8983 | 31.5668 | 30.7429 | 29.3323 |
| ParkScene | 34.3273 | 33.2331 | 31.9147 | 30.3856 | 28.7054 | 34.6558 | 33.6197 | 32.3031 | 30.6253 | 28.8734 |
| Traffic | 37.7826 | 36.0363 | 34.0288 | 31.7721 | 29.3266 | 38.5194 | 36.7086 | 34.5679 | 32.1334 | 29.6176 |

**Mean PSNR (dB) Per Sequence, Per QP: FDPQ Versus RDOQ (YCbCr 4:4:4) – Random Access**

| Sequence | FDPQ | | | | | RDOQ | | | | |
|---|---|---|---|---|---|---|---|---|---|---|
| | QP 17 | QP 22 | QP 27 | QP 32 | QP 37 | QP 17 | QP 22 | QP 27 | QP 32 | QP 37 |
| BirdsInCage | 35.8818 | 34.7279 | 34.3471 | 33.5290 | 32.3055 | 36.5869 | 34.6978 | 34.4153 | 33.6150 | 32.3482 |
| DuckAndLegs | 31.5745 | 29.0506 | 27.8146 | 26.6442 | 25.1170 | 33.3669 | 29.3025 | 27.9705 | 26.8144 | 25.2824 |
| Kimono | 35.1706 | 33.9955 | 33.0732 | 31.7813 | 30.2418 | 35.2126 | 33.9887 | 33.1850 | 31.9445 | 30.4645 |
| OldTownCross | 32.0615 | 30.1570 | 29.9503 | 29.3844 | 28.3097 | 32.7350 | 30.0759 | 30.0171 | 29.4926 | 28.4508 |
| ParkScene | 34.3445 | 32.8848 | 31.7675 | 30.3512 | 28.6899 | 34.6814 | 33.0917 | 32.0614 | 30.6221 | 28.8909 |
| Traffic | 37.6342 | 35.8630 | 33.9318 | 31.7879 | 29.4056 | 38.3907 | 36.5171 | 34.4577 | 32.1694 | 29.7062 |

Overall, the mathematical reconstruction quality of the RDOQ-coded sequences, as quantified by SSIM and PSNR (see Table 4 to Table 7), proved to be superior in the vast majority of cases. However, according to the MOS results obtained via the subjective evaluations, the participants did not notice any perceivable visual quality differences between any of the FDPQ-coded sequences and the RDOQ-coded sequences (see Table 8). This provides evidence that comparatively high SSIM and PSNR values do not necessarily equate to superior perceptual quality. Furthermore, this observation pertains to the fact that subjective evaluations are critically important in terms of robustly assessing HVS-orientated perceptual coding techniques.

**Table 8:** The MOS results, rounded to the nearest integer, of four participants in the subjective evaluations for FDPQ versus RDOQ.

**Rounded Mean Opinion Score (Spatiotemporal Subjective Evaluation) – FDPQ versus RDOQ**

| Sequence | YCbCr 4:2:0 All Intra | | | | YCbCr 4:2:0 Random Access | | | |
|---|---|---|---|---|---|---|---|---|
| | QP 22 | QP 27 | QP 32 | QP 37 | QP 22 | QP 27 | QP 32 | QP 37 |
| BirdsInCage | 5 | 5 | 5 | 5 | 5 | 5 | 5 | 5 |
| DuckAndLegs | 5 | 5 | 5 | 5 | 5 | 5 | 5 | 5 |
| Kimono | 5 | 5 | 5 | 5 | 5 | 5 | 5 | 5 |
| OldTownCross | 5 | 5 | 5 | 5 | 5 | 5 | 5 | 5 |
| ParkScene | 5 | 5 | 5 | 5 | 5 | 5 | 5 | 5 |
| Traffic | 5 | 5 | 5 | 5 | 5 | 5 | 5 | 5 |

**Rounded Mean Opinion Score (Spatiotemporal Subjective Evaluation) – FDPQ versus RDOQ**

| Sequence | YCbCr 4:2:2 All Intra | | | | YCbCr 4:2:2 Random Access | | | |
|---|---|---|---|---|---|---|---|---|
| | QP 22 | QP 27 | QP 32 | QP 37 | QP 22 | QP 27 | QP 32 | QP 37 |
| BirdsInCage | 5 | 5 | 5 | 5 | 5 | 5 | 5 | 5 |
| DuckAndLegs | 5 | 5 | 5 | 5 | 5 | 5 | 5 | 5 |
| Kimono | 5 | 5 | 5 | 5 | 5 | 5 | 5 | 5 |
| OldTownCross | 5 | 5 | 5 | 5 | 5 | 5 | 5 | 5 |
| ParkScene | 5 | 5 | 5 | 5 | 5 | 5 | 5 | 5 |
| Traffic | 5 | 5 | 5 | 5 | 5 | 5 | 5 | 5 |

**Rounded Mean Opinion Score (Spatiotemporal Subjective Evaluation) – FDPQ versus RDOQ**

| Sequence | YCbCr 4:4:4 All Intra | | | | YCbCr 4:4:4 Random Access | | | |
|---|---|---|---|---|---|---|---|---|
| | QP 22 | QP 27 | QP 32 | QP 37 | QP 22 | QP 27 | QP 32 | QP 37 |
| BirdsInCage | 5 | 5 | 5 | 5 | 5 | 5 | 5 | 5 |
| DuckAndLegs | 5 | 5 | 5 | 5 | 5 | 5 | 5 | 5 |
| Kimono | 5 | 5 | 5 | 5 | 5 | 5 | 5 | 5 |
| OldTownCross | 5 | 5 | 5 | 5 | 5 | 5 | 5 | 5 |
| ParkScene | 5 | 5 | 5 | 5 | 5 | 5 | 5 | 5 |
| Traffic | 5 | 5 | 5 | 5 | 5 | 5 | 5 | 5 |

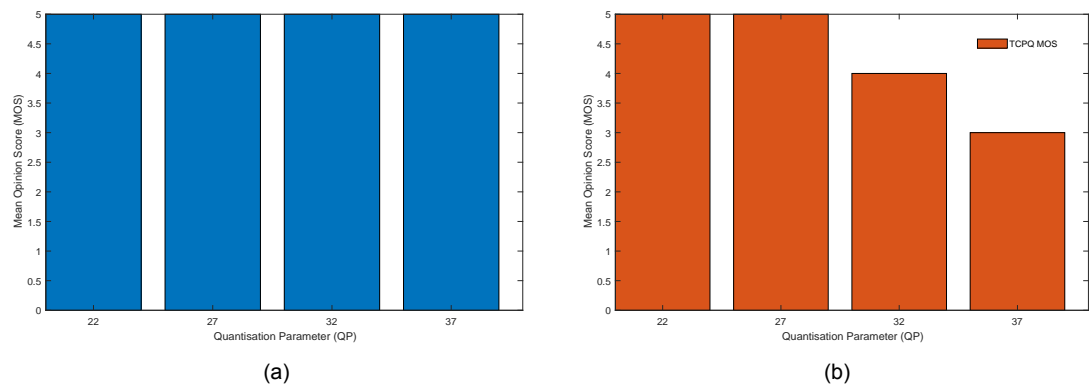

**Figure 15:** Two Mean Opinion Score (MOS) bar graphs. Subfigure (a) shows the MOS for FDPQ versus RDOQ on all sequences using the AI and RA configurations. Subfigure (b) shows the MOS for FDPQ versus the raw video data in the BirdsInCage 4:4:4 10-bit sequence.

**Table 9:** The MOS results, rounded to the nearest integer, of four participants in the subjective evaluations for FDPQ versus the raw video data.

| **Rounded Mean Opinion Score (Spatiotemporal Subjective Evaluation) – FDPQ versus Raw Data** | | | | | | | | |
|---|---|---|---|---|---|---|---|---|
| **Sequence** | YCbCr 4:2:0 All Intra | | | | YCbCr 4:2:0 Random Access | | | |
| | **QP 22** | **QP 27** | **QP 32** | **QP 37** | **QP 22** | **QP 27** | **QP 32** | **QP 37** |
| BirdsInCage | 5 | 4 | 2 | 1 | 5 | 4 | 3 | 2 |
| DuckAndLegs | 5 | 4 | 3 | 1 | 5 | 5 | 4 | 3 |
| Kimono | 5 | 3 | 2 | 1 | 5 | 4 | 3 | 2 |
| OldTownCross | 5 | 3 | 1 | 1 | 5 | 4 | 3 | 2 |
| ParkScene | 5 | 4 | 2 | 1 | 5 | 4 | 3 | 2 |
| Traffic | 5 | 4 | 2 | 1 | 5 | 5 | 4 | 3 |

| **Rounded Mean Opinion Score (Spatiotemporal Subjective Evaluation) – FDPQ versus Raw Data** | | | | | | | | |
|---|---|---|---|---|---|---|---|---|
| **Sequence** | YCbCr 4:2:2 All Intra | | | | YCbCr 4:2:2 Random Access | | | |
| | **QP 22** | **QP 27** | **QP 32** | **QP 37** | **QP 22** | **QP 27** | **QP 32** | **QP 37** |
| BirdsInCage | 5 | 4 | 2 | 1 | 5 | 4 | 3 | 2 |
| DuckAndLegs | 5 | 4 | 3 | 1 | 5 | 5 | 3 | 2 |
| Kimono | 5 | 4 | 2 | 1 | 5 | 4 | 3 | 2 |
| OldTownCross | 5 | 4 | 2 | 1 | 5 | 4 | 3 | 2 |
| ParkScene | 5 | 4 | 2 | 1 | 5 | 5 | 4 | 3 |
| Traffic | 5 | 5 | 2 | 1 | 5 | 5 | 4 | 3 |

| **Rounded Mean Opinion Score (Spatiotemporal Subjective Evaluation) – FDPQ versus Raw Data** | | | | | | | | |
|---|---|---|---|---|---|---|---|---|
| **Sequence** | YCbCr 4:4:4 All Intra | | | | YCbCr 4:4:4 Random Access | | | |
| | **QP 22** | **QP 27** | **QP 32** | **QP 37** | **QP 22** | **QP 27** | **QP 32** | **QP 37** |
| BirdsInCage | 5 | 4 | 3 | 1 | 5 | 5 | 4 | 3 |
| DuckAndLegs | 5 | 5 | 4 | 2 | 5 | 5 | 4 | 3 |
| Kimono | 5 | 4 | 3 | 1 | 5 | 4 | 3 | 2 |
| OldTownCross | 5 | 4 | 2 | 1 | 5 | 4 | 3 | 2 |
| ParkScene | 5 | 4 | 3 | 1 | 5 | 5 | 4 | 3 |
| Traffic | 5 | 4 | 2 | 1 | 5 | 5 | 4 | 3 |

Table 8 and Table 9 tabulate the rounded MOS for the four subjective evaluation participants. Table 8 includes the MOS results for 'FDPQ versus RDOQ' and Table 9 shows the MOS results for 'FDPQ versus the raw video data'. As shown in Table 8, the subjective evaluation participants were unable to detect any perceptually discernible differences between the FDPQ-coded sequences and the RDOQ-coded sequences. An MOS value of 5 is recorded for all tests on all sequences (i.e., the AI and RA tests using initial QPs 17, 22, 27, 32 and 37 on the 4:2:0, 4:2:2 and 4:4:4 versions of each sequence).

The MOS results tabulated in Table 8 are significantly different from those shown in Table 9. Visually lossless coding is achieved by FDPQ in all of the RA QP = 22 tests. Similarly, in almost all of the AI QP = 22 tests, visually lossless coding is accomplished by FDPQ. It is important to note that, in certain QP = 27 tests including the RA QP = 27 tests conducted on the BirdsInCage 4:4:4 and DuckAndLegs 4:4:4 sequences, visually lossless coding is attained by FDPQ; this is significant from a bitrate reduction perspective. Recall from Table 3 (and compared with RDOQ), FDPQ attains 39.4% and 32.8% bitrate reductions when applied to the BirdsInCage 4:4:4 sequence and the DuckAndLegs 4:4:4 sequence, respectively. In the vast majority of cases, the FDPQ-coded sequences using the RA encoding configuration — i.e., the RA Group Of Pictures (GOP)-based inter coding tests — were perceived to be vastly superior compared with the sequences coded using the All Intra configuration. This is because motion data with GOP-based inter coding in HEVC can be signalled with the utilisation of merge mode or by motion vector differences, picture reference indices and the direction of the inter prediction [19, 20].

Recall that the 4:4:4 (and 4:2:2) versions of the BirdsInCage sequence and the DuckAndLegs sequence are high bit depth 30-bit sequences (i.e., 10-bits per channel). Note that 30-bit video data contains a much larger number of colours per pixel compared with 24-bit video data (i.e., potentially up to $1024^3$ colours per pixel). Therefore, in combination with the absence of chroma subsampling in YCbCr 4:4:4 data, the high bit depth characteristics of these 30-bit sequences equates to the fact that the 10-bit Y, Cb and Cr channels comprise higher variances compared with the 8-bit Y, Cb and Cr channels in 24-bit YCbCr 4:2:0 chroma subsampled video data. To reiterate, it has been established in the literature that it is more difficult for the HVS to detect compression-induced artifacts in high variance regions of image and video data, which constitutes high variance-based spatial masking [21]-[23]. As shown in other previously published work, this visual masking phenomenon is more prominent in high bit depth 4:4:4 data [24, 25]. With this in mind, discarding high frequency detail in high variance luma and chroma data (i.e., 30-bit YCbCr 4:4:4 sequences) is typically not noticeable to the HVS.

In relation to the exhaustive experimental evaluations conducted in this paper, it is evident that visually lossless coding can be achieved without utilising complex psychovisual models in the frequency domain. Although FDPQ is a HVS-orientated technique that consists of perceptual considerations (i.e., quantising high frequency AC coefficient more coarsely than the DC coefficient and the low frequency AC coefficient), FDPQ is based on a conceptually simple adaptive Euclidean distance parameter. It is important to note that both FDPQ and RDOQ are not suitable for low bitrate All Intra coding applications, as confirmed in the subjective evaluation results for the AI QP = 37 tests (see Table 9). This is due to the fact that both FDPQ and RDOQ are designed, for the most part, to preserve the integrity of the DC transform coefficient and low frequency AC coefficients in luma and chroma TBs.

In accordance with the subjective evaluation results, the quantisation-induced compression artifacts incurred by FDPQ in the RA QP = 37 tests are considerably less conspicuous than those that were induced in AI QP = 37 tests. To reiterate, this is because GOP-based inter coding includes the signalling of important motion data in the bitstream; All Intra coding does not account for motion data or the spatiotemporal redundancies that exist between frames. Therefore, the visual quality of the reconstructed inter-coded sequences — for both FDPQ and RDOQ — is significantly superior compared with the corresponding intra-coded sequences.

**5.0 Conclusion**

In this paper, a novel coefficient-level perceptual quantisation technique, named FDPQ, has been proposed. FDPQ applies coarse perceptual quantisation to high frequency AC coefficients in the frequency domain by exploiting the MTF characteristics of the HVS. By utilising a novel distance parameter, FDPQ measures the distance of an AC coefficient from the DC coefficient and subsequently discards the perceptually insignificant AC coefficients from the high frequency sub-band in the frequency domain. Compared with the ubiquitous coefficient-level quantisation technique known as RDOQ, FDPQ attains considerable bitrate reductions of 41% without incurring a perceptually conspicuous decrease of visual quality in the compressed video data. The experimental evaluation results, in both the objective and subjective evaluations, highlight the fact that FDPQ is vastly more effective on high bit depth and full chroma sampling video data (i.e., YCbCr 4:4:4 10-bit data). With relevance to perceptual quantisation applications, this observation may give rise to new lines of research as regards the potential importance of chroma sampling and the bit depth of raw YCbCr data. In terms of runtimes, no significant differences are observed; slight reductions in encoding times and decoding times are achieved by FDPQ.

**Acknowledgements**

I wish to show gratitude to Dr. Victor Sanchez for the valuable feedback provided regarding the contributions presented in this paper. Dr. Victor Sanchez presently holds an associate professorship tenure in the Department of Computer Science at the University of Warwick, England, UK.